\documentclass[12pt]{iopart}
\usepackage{graphicx}
\usepackage{amsthm}
\usepackage{cite}
\newtheorem{theorem}{Theorem}[section]
\newtheorem{lemma}[theorem]{Lemma}

\newenvironment{definition}[1][Definition]{\begin{trivlist}
\item[\hskip \labelsep {\bfseries #1}]}{\end{trivlist}}

\begin{document}

\title[Robustness of Random Graphs]{Robustness of Random Graphs Based on Natural Connectivity}

\author{Jun Wu$^{1,2}$, Mauricio Barahona$^{2,3}$, Yuejin Tan$^1$ and Hongzhong Deng$^1$}

\address{$^1$ College of Information Systems and Management, National University of Defense Technology, Changsha 410073, P. R. China}
\address{$^2$ Institute for Mathematical Sciences, Imperial College London, London SW7 2PG, United Kingdom}
\address{$^3$ Department of Bioengineering, Imperial College London, London SW7 2AZ, United Kingdom}
\ead{wujunpla@hotmail.com}

\begin{abstract}
Recently, it has been proposed that the natural connectivity can be used to efficiently characterise the robustness
of complex networks. Natural connectivity quantifies the redundancy of alternative routes in a network by evaluating the weighted number of closed walks of all lengths and can be regarded as the average eigenvalue obtained from the graph spectrum. In this article, we explore the natural connectivity of random graphs both analytically and numerically and show that it increases linearly with the average degree. By comparing with regular ring lattices and random regular graphs, we show that random graphs are more robust than random regular graphs; however, the relationship between random graphs and regular ring lattices depends on the average degree and graph size. We derive the critical graph size as a function of the average degree, which can be predicted by our analytical results. When the graph size is less than the critical value, random graphs are more robust than regular ring lattices, whereas regular ring lattices are more robust than random graphs when the graph size is greater than the critical value.
\end{abstract}

%Uncomment for PACS numbers title message
%\pacs{00.00, 20.00, 42.10}
% Keywords required only for MST, PB, PMB, PM, JOA, JOB?
%\vspace{2pc}
%\noindent{\it Keywords}: Article preparation, IOP journals
% Uncomment for Submitted to journal title message
%\submitto{\JPA}
% Comment out if separate title page not required
\maketitle

\section{Introduction}
Networks are everywhere. Many systems in nature and society can be described as complex networks. Examples of networks include the Internet \cite{Vazquez}, metabolic networks~\cite{Jeong}, electric power grids~\cite{Rosas-Casals}, supply chains~\cite{Thadakamalla}, urban road networks~\cite{Xie}, world trade web~\cite{Serrano} and many others. Complex networks, more generally, complex systems have become pervasive in today's science and technology scenario and have recently become one of the most popular topics within the interdisciplinary area involving physics, mathematics, biology, social sciences, informatics, and other theoretical and applied sciences(see~\cite{Albert1,Newman1,Amaral}). Complex networks rely for their function and performance on their robustness, that is, the ability to endure threats and survive accidental events. Due to their broad range of applications, the attack robustness of complex networks has received growing attention.

Simple and effective measures of robustness are essential for the study of robustness. A variety of measures, based on different heuristics, have been proposed to quantify the robustness of networks. For instance, the vertex (edge) connectivity of a graph is an important, and probably the earliest, measure of robustness of a network~\cite{Whitney}. However, the vertex (edge) connectivity only partly reflects the ability of graphs to retain connectedness after vertex (or edge) deletion. Other improved measures include super connectivity~\cite{Bauer}, conditional connectivity~\cite{Harary}, restricted connectivity~\cite{Esfahanian}, fault diameter~\cite{Krishnamoorthy}, toughness~\cite{Chvatal}, scattering number~\cite{Jung}, tenacity~\cite{Cozzen}, the expansion parameter~\cite{Alon} and the isoperimetric number~\cite{Mohar}. In contrast to vertex (edge) connectivity, these new measures consider both the cost to damage a network and how badly the network is damaged. Unfortunately, from an algorithmic point of view, the problem of calculating these measures for general graphs is NP-complete. This implies that these measures are of no great practical use within the context of complex networks. Another remarkable measure used to unfold the robustness of a network is the second smallest (first non-zero) eigenvalue of the Laplacian matrix, also known as the algebraic connectivity. Fiedler~\cite{Fiedler} showed that the magnitude of the algebraic connectivity reflects how well connected the overall graph is; the larger the algebraic connectivity is, the more difficult it is to cut a graph into independent components. However, the algebraic connectivity is equal to zero for all disconnected networks. Therefore, it is too coarse a measure to be used for complex networks..

An alternative formulation of robustness within the context of complex networks emerged from the random graph theory~\cite{Bollobas} and was stimulated by the work of Albert et al.~\cite{Albert2}. Instead of a strict extreme property, they proposed a statistical measure, that is, the critical removal fraction of vertices (edges) for the disintegration of a network, to characterise the robustness of complex networks. The disintegration of networks can be observed from the decrease of network performance. The most common performance measurements include the diameter, the size of the largest component, the average path length, the efficiency~\cite{Latora}  and the number of reachable vertex pairs~\cite{Siganos}. As the fraction of removed vertices (or edges) increases, the performance of the network will eventually collapse at a critical fraction. Although we can obtain the analytical critical removal fraction for some special networks~\cite{Cohen1,Cohen2,Callaway,Wu1,Wu2}, generally, this measure can only be calculated through simulations.

Recently, Wu {\it et al.}~\cite{Wu3} showed that the concept of natural connectivity can be used to characterize the robustness of complex networks. The concept of natural connectivity is based on the Estrada index of a graph, which has been proposed to characterize molecular structure~\cite{Estrada2000}, bipartivity~\cite{Estrada2005a}, subgraph centrality~\cite{Estrada2005b} and expansibility~\cite{Estrada2006a, Estrada2006b}. Natural connectivity has an intuitive physical meaning and a simple mathematical formulation. Physically, it characterises the redundancy of alternative paths by quantifying the weighted number of closed walks of all lengths leading to a measure that works in both connected and disconnected graphs. Mathematically, the natural connectivity is obtained from the graph spectrum as an average eigenvalue and it increases strictly monotonically with the addition of edges. Abundant information about the topology and dynamical processes can be extracted from a spectral analysis of the networks. Natural connectivity sets up a bridge between the graph spectra and the robustness of complex networks, allowing a precise quantitative analysis of the network robustness.

In our previous study~\cite{Wu4}, we have shown that the natural connectivity of regular ring lattices is independent of the network size and increases linearly with the average degree. In this study, we investigate the natural connectivity of random graphs and compare it with regular graphs. The article is structured as follows. In Section 2, we introduce the concept of natural connectivity and some basic elements of random graphs. In Section 3, we derive the natural connectivity of random graphs. In Section 4, we compare the natural connectivity of random graphs with that of regular graphs. Finally, the conclusions are presented in Section 5.

\section{Preliminaries}
\subsection{Graph and Natural Connectivity}

A complex network can be viewed as a simple undirected graph $G(V,E)$, where $V$ is the set of vertices, and $E \subseteq V \times V$ is the set of edges. Let $N = \left| V \right|$ and $M = \left| E \right|$ be the number of vertices and the number of edges, respectively. Let $A(G) = (a_{ij} )_{N \times N} $ be the adjacency matrix of $G$, where $a_{ij}  = a_{ji}  = 1$ if vertices $v_i $ and $v_j $ are adjacent, and $a_{ij}  = a_{ji}  = 0$ otherwise. It is obvious that $A(G)$ is a real symmetric matrix. We thus let $\lambda _1  \ge \lambda _2  \ge ... \ge \lambda _N $ denote the eigenvalues of $A$ which are usually also called the eigenvalues of the graph $G$ itself. The set $\left\{ {\lambda _1 ,\lambda _2 ,...\lambda _N } \right\}$ is called the spectrum of $G$. The spectral density of $G$ is defined as the sum of the $\delta $ functions as follows
\begin{equation}
\label{spectral_density}
\rho (\lambda ) = \frac{1}{N}\sum\limits_{i = 1}^N {\delta {\rm{(}}\lambda {\rm{ - }}\lambda _i {\rm{)}}}
\end{equation}
which converges to a continuous function when $N \to \infty $, where $\delta {\rm{(}}\lambda {\rm{ - }}\lambda _i {\rm{) = 1}}$ if $\lambda {\rm{ = }}\lambda _i$; and $\delta {\rm{(}}\lambda {\rm{ - }}\lambda _i {\rm{) = 0}}$, otherwise.

A walk of length $k$ in a graph $G$ is an alternating sequence of vertices and edges $v_0 e_1 v_1 e_2 ...e_k v_k $, where $v_i  \in V$ and $e_i  = (v_{i - 1} ,v_i ) \in E$. A walk is closed if $v_0  = v_k $. The number of closed walks is an important index for complex networks. Recently, we have defined the redundancy of alternative paths as the number of closed walks of all lengths~\cite{Wu3}. Considering that shorter closed walks have more influence on the redundancy of alternative paths than longer closed walks and to avoid the number of closed walks of all lengths to diverge, we scale the contribution of closed walks to the redundancy of alternative paths by dividing them by the factorial of the length k. That is, we define a weighted sum of numbers of closed walks $S = \sum\nolimits_{k = 0}^\infty  {n_k /k!}$, where $n_k $ is the number of closed walks of length $k$. This scaling ensures that the weighted sum does not diverge and it also means that S can be obtained from the powers of the adjacency matrix:
\begin{equation}
\label{n_k}
n_k  = trace(A^k ) = \sum\limits_{i = 1}^N {\lambda _i ^k }
\end{equation}
Using Eq.~\ref{n_k}, we can obtain
\begin{equation}\label{S}
S = \sum\limits_{k = 0}^\infty  {\frac{{n_k }}{{k!}}}  = \sum\limits_{k = 0}^\infty  {\sum\limits_{i = 1}^N {\frac{{\lambda _i ^k }}{{k!}}} }  = \sum\limits_{i = 1}^N {\sum\limits_{k = 0}^\infty  {\frac{{\lambda _i ^k }}{{k!}}} }  = \sum\limits_{i = 1}^N {e^{\lambda _i } } .
\end{equation}

Hence the proposed weighted sum of closed walks of all lengths can be derived from the graph spectrum. We remark that Eq.~\ref{S} corresponds to the Estrada Index of the graph~\cite{Estrada2000}, which has been used in several contexts in the graph theory, including bipartivity~\cite{Estrada2005a} and subgraph centrality~\cite{Estrada2005b}. The natural connectivity can be defined as the average eigenvalue of the graph as follows.
\begin{definition}~\cite{Wu3}
Let $A(G)$ be the adjacency matrix of $G$ and let $\lambda _1  \ge \lambda _2  \ge ... \ge \lambda _N $ be the eigenvalues of $A(G)$. Then the natural connectivity or natural eigenvalue of $G$ is defined by
\begin{equation}\label{nat}
\bar \lambda  = \ln \left( {\sum\limits_{i = 1}^N {e^{\lambda _i } } /N} \right)
\end{equation}
\end{definition}

It is evident from Eq.~\ref{nat} that $\lambda _N  \le \bar \lambda  \le \lambda _1 $.

A regular ring lattice $RRL_{N,2K} $ is a $2K - regular $ graph with $N$ vertices on a one-dimensional lattice, in which each vertex is connected to its $2K$ neighbors ($K$ on either side). In a previous study~\cite{Wu4}, we have investigated the natural connectivity of regular ring lattices and shown that random regular graphs are less robust than regular ring lattices based on natural connectivity.
\begin{theorem}~\cite{Wu4} Let $RRL_{N,2K} $ be a regular ring lattice.Then the natural connectivity of $RRL_{N,2K} $ is
\begin{equation}\label{nat_rrl}
\bar \lambda _{R_{N,2K} }  = \ln \left( {I_0 (\overbrace {2,2,...2}^K)} \right){\rm{  + }}o{\rm{(1) }}
\end{equation}
\end{theorem}
where $o{\rm{(1)}} \to {\rm{0 }}$ as $N \to \infty$.

\subsection{Erd\H os-R\'enyi Random Graphs}

Random graphs have long been used for modelling the topology of systems made up of large assemblies of similar units. The theory of random graphs was introduced by Erd\H{o}s {\it et al.}~\cite{Erdos1}. A detailed review of random graphs can be found in the classic book~\cite{Bollobas}. A random graph is obtained by starting with a set of $N$ vertices and adding edges between them at random. In this article, we study the random graphs of the classic Erd\H os-R\'enyi model $G_{N,p} $, in which each of the possible $C_N^2  = N(N - 1)/2$  edges occurs independently with probability $p$. Consequently, the total number of edges $M$ is a random variable with the expectation value $E(M) = p \cdot C_N^2 $ and then the average degree $ < k >  = (N - 1)p \approx Np$.

Random graph theory studies the properties of the probability space associated with graphs with $N$ vertices as $N \to \infty $. Many properties of such random graphs can be determined using probabilistic arguments. We say that a graph property $Q$ holds almost surely for $G_{N,p} $ if the probability that $G_{N,p} $ has property $Q$ tends to one as $N \to \infty $. Furthermore, Erd\H{o}s {\it et al.} described the behavior of $G_{N,p} $ very precisely for various values of $p$ ~\cite{Erdos2}. Their results showed that:
\begin{enumerate}
\item
If $Np < 1$, then a graph $G_{N,p} $ will almost surely have no connected components of size larger than $o(\ln N)$; If $Np \ge 1$, then a graph $G_{N,p} $ will almost surely have a unique "giant" component containing a positive fraction of the vertices.
\item
If $Np < \ln N$, then a graph $G_{N,p} $ will almost surely not be connected; If $Np \ge \ln N$, then a graph $G_{N,p} $ will almost surely be connected.
\end{enumerate}

It is well known that the largest eigenvalue $\lambda _1 $ of $G_{N,p} $ is almost surely $Np[1 + o(1)]$ provided that $Np \gg \ln N$(see~\cite{Chung,Krivelevich}). Moreover, according to the famous Wigner's law or semicircle law~\cite{Wigner}, as $N \to \infty $, the spectral density of $G_{N,p} $ converges to a semicircular distribution as follows
\begin{equation}
\rho (\lambda ) = \left\{ {\begin{array}{*{20}c}
   {\frac{{2\sqrt {R^2  - \lambda ^2 } }}{{\pi R^2 }}} & {\left| \lambda  \right| \le R}  \\
   0 & {\left| \lambda  \right| > R}  \\
\end{array}} \right.
\end{equation}
where $R = 2\sqrt {Np(1 - p)} $ is the radius of the "bulk" part of the spectrum.

\section{Natural connectivity of random graphs}

When $N \to \infty $, with continuous approximation for $\lambda _i $, Eq.~\ref{nat} can be rewritten in the following spectral density form
\begin{equation}
\bar \lambda  = \ln \left( {\int_{ - \infty }^{ + \infty } {\rho (\lambda )e^\lambda  } d\lambda } \right) = \ln \left( {M_\lambda  (1)} \right)
\end{equation}
where $\rho (\lambda )$ is the spectral density and $M_\lambda  (t)$ is the moment generating function of $\rho (\lambda )$. Consequently, we obtain the natural connectivity of Erd\H{o}s-R\'enyi random graphs with $p \gg \ln N/N$
\begin{equation}
\bar \lambda  = \ln \left( {\int_{ - R}^{ + R} {\rho (\lambda )e^\lambda  } d\lambda  + e^{\lambda _1 } /N} \right) = \ln \left( {M_\lambda  (1) + e^{Np} /N} \right)
\end{equation}
where
\begin{equation}
M_\lambda  (1) = \int_{ - R}^{ + R} {\frac{{2\sqrt {R^2  - \lambda ^2 } }}{{\pi R^2 }}e^\lambda  } d\lambda  = \frac{2}{\pi }\int_{ - R}^{ + R} {\frac{{\sqrt {R^2  - \lambda ^2 } }}{{R^2 }}e^\lambda  d\lambda }
\end{equation}
Substituting $\lambda  = R\cos (\theta )$ into Eq. (8), we obtain that
\begin{equation}
M_\lambda  (1) = \frac{2}{\pi }\int_0^\pi  {e^{R\cos (\theta )} } \sin ^2 (\theta )d\theta
\end{equation}
Note that~\cite{Abramowitz}
\begin{equation}
I_\alpha  (x) = \frac{{(x/2)^\alpha  }}{{\pi ^{1/2} \Gamma (\alpha  + 1/2)}}\int_0^\pi  {e^{x\cos (\theta )} } \sin ^{2\alpha } (\theta )d\theta
\end{equation}
where $I_\alpha  (x)$ is the modified Bessel function and $\Gamma (x)$ is the Gamma function. Then we obtain that
\begin{equation}
I_1 (R) = \frac{R}{\pi }\int_0^\pi  {e^{R\cos (\theta )} } \sin ^2 (\theta )d\theta
\end{equation}
Using Eq. (11), we can simplify Eq. (9) as
\begin{equation}
M_\lambda  (1) = 2I_1 (R)/R
\end{equation}
Substituting Eq. (12) into Eq. (7), we obtain that
\begin{equation}
\bar \lambda  = \ln \left( {\frac{2{I_1 (R)}}{R} + \frac{{e^{Np} }}{N}} \right) = Np - \ln (N) + \ln \left( {1 + \frac{2{NI_1 (R)}}{{e^{Np} R}}} \right)
\end{equation}

Now we propose two lemmas first.
\begin{lemma}
As $N \to \infty $, $f(p) = 2NI_1 (R)/(e^{Np} R)$ is a monotonically decreasing function for $\ln N/N < p < 1 - \ln N/N$, where $R = 2\sqrt {Np(1 - p)} $.
\end{lemma}
\proof
It is easy to know that $2\sqrt {\ln N(1 - \ln N/N)}  < R \le \sqrt N $ for $\ln N/N < p < 1 - \ln N/N$. Then as $N \to \infty $, we have $R \to \infty $. Note that, for the large values of $x \gg \left| {\alpha ^2  - 1/4} \right|$, the modified Bessel functions $I_\alpha  (x)$ have the following asymptotic forms~\cite{Arfken}
\begin{equation}
I_\alpha  (x) \to \frac{1}{{\sqrt {2\pi x} }}e^x
\end{equation}
Thus, for $\ln N/N < p < 1 - \ln N/N$, we obtain
\begin{equation}
I_1 (R) \to \frac{1}{{\sqrt {2\pi R} }}e^R
\end{equation}
Then we have
\begin{equation}
f(p) \to N\sqrt {\frac{2}{\pi }}  \cdot \frac{{e{}^{R - Np}}}{{R^{3/2} }}
\end{equation}
Note that,
\begin{equation}
\begin{array}{c}
 \frac{{df(p)}}{{dp}} \to \frac{{e^{R - Np} \left( {\frac{{dR}}{{dp}} - N} \right)R^{3/2}  - \frac{3}{2}R^{1/2}  \cdot \frac{{dR}}{{dp}} \cdot e^{R - Np} }}{{R^3 }} \\
  = \frac{N}{{R^3 }}\sqrt {\frac{2}{\pi }} \left( {N(2 - 4p - R) - \frac{{3N(1 - 2p)}}{R}} \right) < 0 \\
 \end{array}
\end{equation}
Therefore, we prove that, as $N \to \infty $, $f(p)$ is a monotonically decreasing function for $\ln N/N < p < 1 - \ln N/N$.
\endproof

\begin{lemma}
Let $p_c  = N^{\varepsilon  - 1} \ln N$, where $0 < \varepsilon  \ll 1$. Then we have $f(p_c ) \to 0$ as $N \to \infty $.
\end{lemma}
\proof
Note that $0 < \varepsilon  \ll 1$, thus we have $p_c  \to 0$ and $1 - p_c  \to 1$ as $N \to \infty $. Then we obtain that $R_{p_c }  \to 2\sqrt {N^\varepsilon  \ln N} $. Therefore, we have
\begin{equation}
\begin{array}{l}
 f(p_c ) \to N\sqrt {\frac{2}{\pi }}  \cdot \frac{{e{}^{R_{p_c }  - Np_c }}}{{R_{p_c } ^{3/2} }} = N\sqrt {\frac{2}{\pi }}  \cdot \frac{{e{}^{2\sqrt {N^\varepsilon  \ln N}  - N^\varepsilon  \ln N}}}{{\left( {2\sqrt {N^\varepsilon  \ln N} } \right)^{3/2} }} \\
  = \frac{{N^{1 - 3\varepsilon /4} }}{{2\sqrt \pi  }} \cdot \frac{{e{}^{2\sqrt {N^\varepsilon  \ln N}  - N^\varepsilon  \ln N}}}{{\left( {\ln N} \right)^{3/4} }} = \frac{{N^{1 - 3\varepsilon /4} }}{{2\sqrt \pi  }} \cdot \frac{{e{}^{2\sqrt {N^\varepsilon  \ln N}  - N^\varepsilon  \ln N}}}{{\left( {\ln N} \right)^{3/4} }} \\
  = \frac{{N^{1 - 3\varepsilon /4} }}{{2\sqrt \pi  }} \cdot \frac{{e{}^{ - \left( {\sqrt {N^\varepsilon  \ln N}  - 1} \right)^2  + 1}}}{{\left( {\ln N} \right)^{3/4} }} \\
 \end{array}
\end{equation}
Since $\sqrt {N^\varepsilon  \ln N}  \gg 1$ as $N \to \infty $, we obtain that
\begin{equation}
f(p_c ) \to \frac{{N^{1 - 3\varepsilon /4} }}{{2\sqrt \pi  }} \cdot \frac{{e{}^{ - \left( {\sqrt {N^\varepsilon  \ln N}  - 1} \right)^2  + 1}}}{{\left( {\ln N} \right)^{3/4} }} \approx \frac{{eN^{1 - 3\varepsilon /4 - N^\varepsilon  } }}{{2\sqrt \pi  \left( {\ln N} \right)^{3/4} }} \to 0
\end{equation}
The proof is completed.
\endproof

From Lemmas 3.1 and 3.2, it is easy to derive that, for $p_c  \le p \le 1 - p_c $, $f(p) \le f(p_c ) \to 0$  as $N \to \infty $. Consequently, we obtain the following theorem.
\begin{theorem}
Let $G_{N,p} $ be a random graph with $N^{\varepsilon  - 1} \ln N < p < 1 - N^{\varepsilon  - 1} \ln N$, where $0 < \varepsilon  \ll 1$. Then the natural connectivity of $G_{N,p} $ almost surely is
\begin{equation}
\bar \lambda  = Np - \ln (N) + o(1)
\end{equation}
where $o{\rm{(1)}} \to {\rm{0 }}$ as $N \to \infty$.
\end{theorem}

From Eq. (20), we know that the natural connectivity of random graphs increases linearly with edge density $p$ given the graph size $N$. Note that $ < k >  = Np$; thus, we also observe that the natural connectivity of random graphs increases linearly  with the average degree given the graph size $N$. To verify our result, we simulate 1000 independent $G_{N,p} $ and compute the average natural connectivity
for each combination of $N$ and $p$. In Fig. 1, we plot the natural connectivity of random graphs with both numerical results and analytical results. We observe that the numerical results agree well with the analytical results.

\begin{figure}[htbp]
\begin{center}
\includegraphics[width=0.8\textwidth]{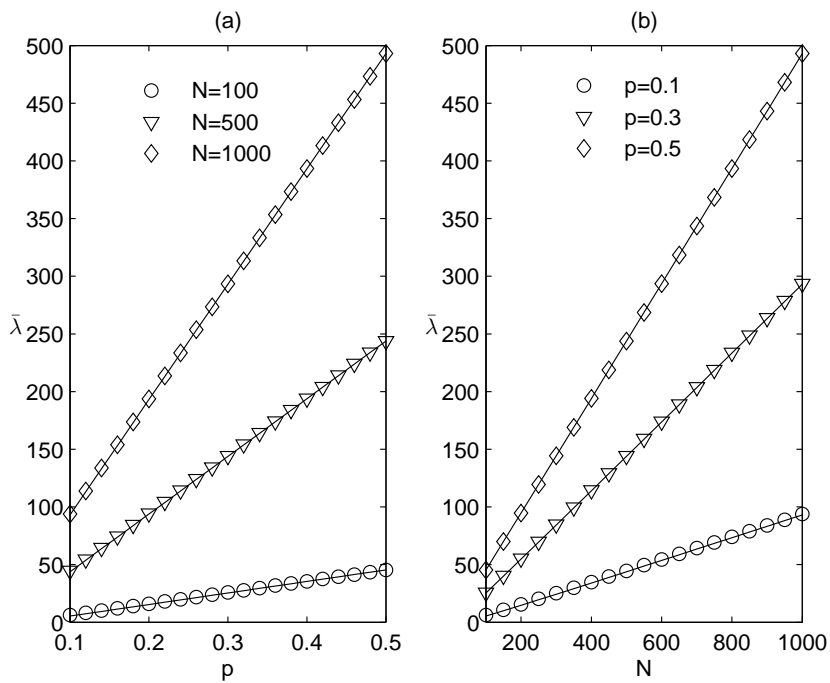}
\caption{Natural connectivity of random graphs: (a) $\bar \lambda $ vs. $p$ with $N = 100$ (circles), 500 (triangles) and 1000 (diamonds); (b) $\bar \lambda $ vs. $N$ with $p = 0.1$ (circles), 0.3 (triangles), 0.5 (diamonds). Each quantity is an average over 1000 realizations. The lines represent the corresponding analytical results according to Eq. (20).}
\end{center}
\end{figure}

\section{Comparisons between random graphs and regular graphs}

We generate Erd\H os-R\'enyi random graphs $G_{N,p} $, regular ring lattices $RRL_{N,2K} $, and random regular graphs $RRG_{N,2K} $ using the algorithm in~\cite{Steger}. We compare the natural connectivity of random graphs with  two other types of regular graphs with the same number of vertices and edges, i.e., $p \approx 2K/N$ and $ < k >  = 2K$. The results are shown in Fig . 2. We find that regular ring lattices and random graphs are always robustness than random regular graphs. However, the curves of regular ring lattices cross those of random graphs; furthermore, random graphs are more robust than regular ring lattices prior to crossings (dense networks), whereas regular ring lattices are more robust than random graphs over crossings (sparse networks). This means that there is a critical graph size $N_c $, that is  as a function of $K$. For example, for $K = 5$, we find that $N_c  \approx 60$.

\begin{figure}[htbp]
\begin{center}
\includegraphics[width=0.8\textwidth]{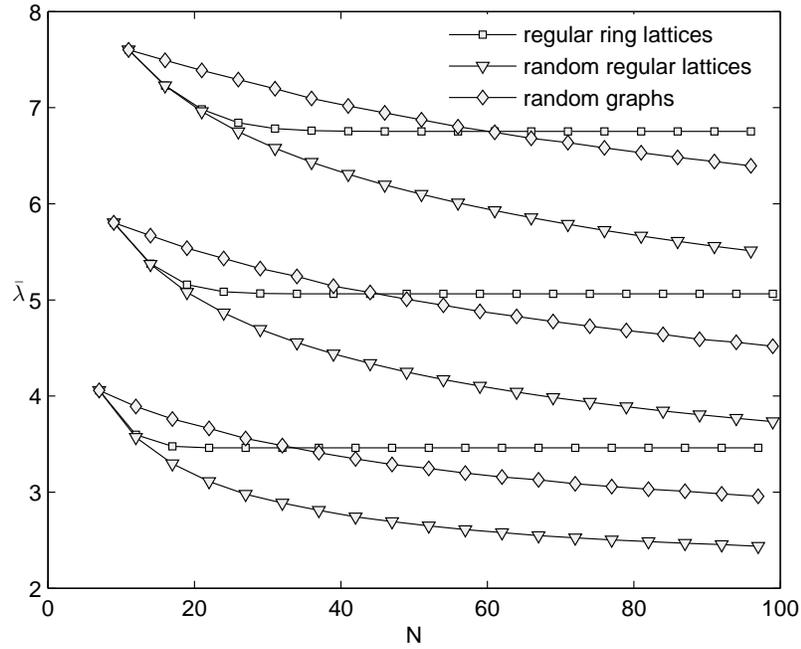}
\caption{Natural connectivity of random graphs of regular ring lattices $RRL_{N,2K} $ (squares), random regular graphs $RRG_{N,2K} $ (triangles) and random graphs $G_{N,p} $ (diamonds) with the same number of vertices and edges. From bottom to top, the symbols correspond to $K = 3,4,5$, respectively. Each quantity is an average over 1000 realizations.}
\end{center}
\end{figure}

For large values of $K$, we can analytically predict the values of $N_c $ using Eq. (4) and Eq. (20) as follows
\begin{equation}
\ln \left( {I_0 (\overbrace {2,2,...2}^K)} \right) = Np - \ln (N) = 2K - \ln (N) \Rightarrow N_c  \approx e^{2K - I_0 (\overbrace {2,2,...2}^K)}
\end{equation}
The results are shown in Fig. 3. Moreover, we also find that there is a critical value $p_c $ or $K_c $ as a function of graph size $N$. Regular ring lattices are more robust than random graphs when the edge density $p < p_c $, whereas random graphs are more robust than regular ring lattices when the edge density $p > p_c $.

\begin{figure}[htbp]
\begin{center}
\includegraphics[width=0.8\textwidth]{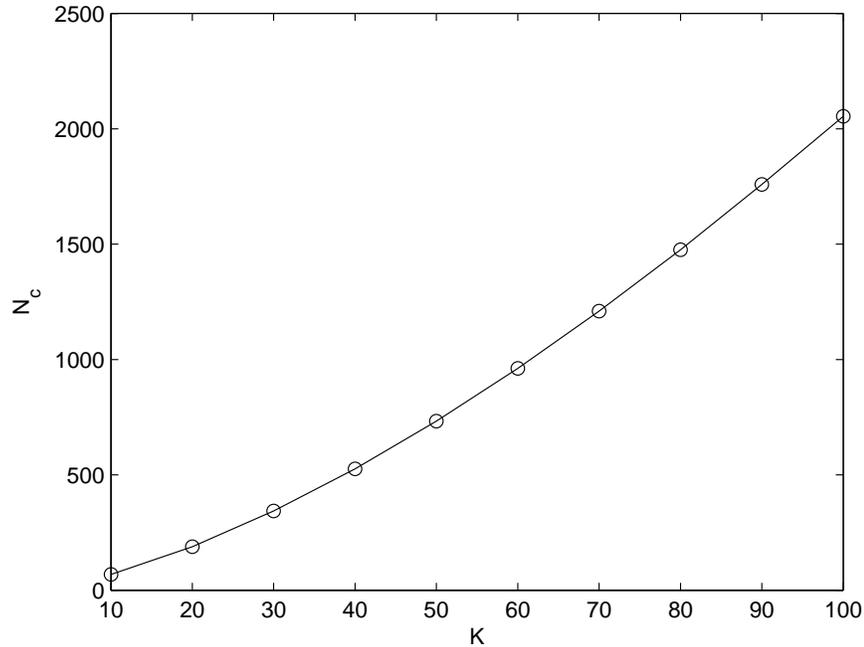}
\caption{The critical value $N_c $ as a function of graph size $K$ according to Eq. (21).}
\end{center}
\end{figure}

To explore the critical behaviors of graphs in depth, we randomise regular ring lattices by random rewiring~\cite{Watts} and by random degree-preserving rewiring~\cite{Newman2}, which leads to random graphs and random regular graphs, respectively. In Fig. 4, the natural connectivity is represented as a function of the number of rewirings, starting from regular ring lattices with $N = 30 < N_c $ and $N = 100 > N_c $, where $K = 5$. We find that the natural connectivity decreases during the process of random degree-preserving rewiring and equals to the value of a random regular graph finally. It means that regular ring lattices are more robust than random regular graphs for both $N = 30$ and $N = 100$. The case of random rewiring is more complicated. Different processes of random rewiring for $N = 30$ and $N = 100$ are shown in Fig. 4. The natural connectivity increases during the process of random rewiring for $N = 30 < N_c $; however, for $N = 100 > N_c $, the natural connectivity first decreases during the process of random rewiring and then increases during the process of random rewiring; finally, equals to the value of a random graph finally (smaller than the value of a regular ring lattice). It means that randomness increases the robustness of a dense regular ring lattice, but decreases the robustness of a sparse regular ring lattice.

\begin{figure}[htb]
\begin{center}
\includegraphics[width=0.8\textwidth]{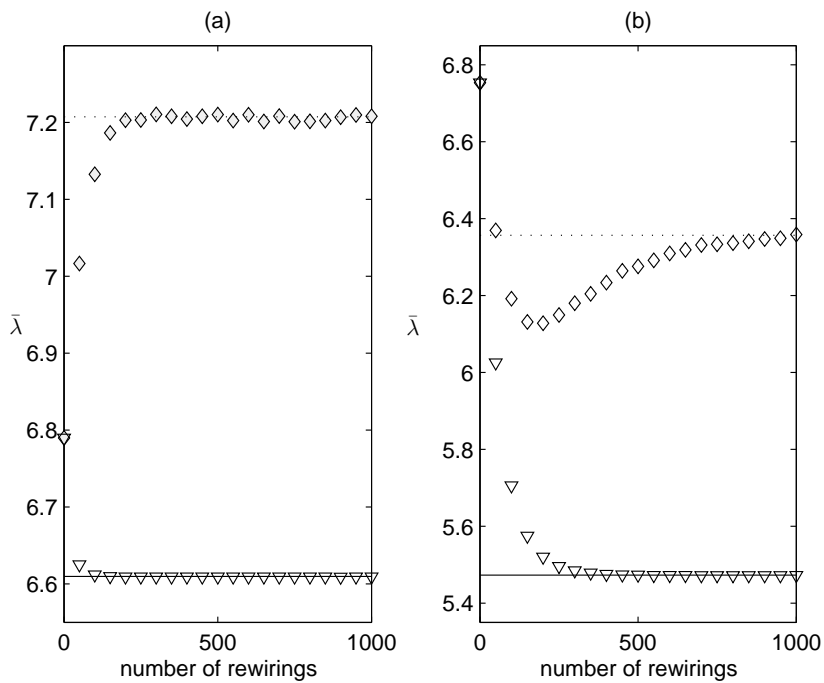}
\caption{Natural connectivity during the processes of random rewiring (diamonds) and random degree-preserving rewiring (triangles) starting from regular ring lattices with $N = 30$ (a), $N = 100$ (b), where $K = 5$. The solid lines represent the values of random regular graphs and the dashed lines represent the values of random graphs. Each quantity is an average over 1000 realizations.}
\end{center}
\end{figure}

\section{Conclusions}

We have investigated the natural connectivity of Erd\H os-R\'enyi random graphs $G_{N,p} $ in this article. We have presented the spectral density form of natural connectivity and derived the natural connectivity of random graphs analytically using the Wigner's semicircle law. In addition, we have shown that the natural connectivity of random graphs increases linearly with edge density $p$ given a large graph size $N$. The analytical results agree with the numerical results very well.

We have compared the natural connectivity of random graphs $G_{N,p} $ with regular ring lattices $RRL_{N,2K} $ and random regular graphs $RRG_{N,2K} $ with the same number of vertices and edges. We have shown that random graphs are more robust than random regular graphs; however the relationship between random graphs and regular ring lattices depends on the graph size $N$ and the edge density $p$ or the average degree $ < k > $. We have observed that the critical value $N_c $ as a function of $K$, and the critical value $p_c $ and $K_c $ as a function of graph size $N$, which can be predicted by our analytical results. We have explored the critical behavior by random rewiring from regular ring lattices. We have shown that randomness increases the robustness of a dense regular ring lattice, but decreases the robustness of a sparse regular ring lattice. Our results will be of great theoretical and practical significance to the network robustness design and optimization.

\ack{This work is in part supported by the National Science
Foundation of China under grant nos 60904065 and 70771111, and the Specialized Research Fund for the Doctoral Program of Higher Education under Grant No 20094307120001.}

\section*{References}
\bibliographystyle{iopart-num}
\bibliography{wu}

\end{document}